\documentclass[12pt]{article}
\usepackage{graphicx}

\newcommand\pubnumber{DPF2013-35}
\newcommand\pubdate{\today}

\def\kennesaw{Kennesaw State University, Kennesaw, GA 30144, USA}
\def\support{\footnote{This material is based upon work supported by the National Science Foundation under Grant No. PHY 1212472.}}

\def\Title#1{\begin{center} {\Large #1 } \end{center}}
\def\Author#1{\begin{center}{ \sc #1} \end{center}}
\def\Address#1{\begin{center}{ \it #1} \end{center}}

\newcommand\pubblock{\rightline{\begin{tabular}{l} \pubnumber\\
         \pubdate  \end{tabular}}}
\newenvironment{Abstract}{\begin{quotation}  }{\end{quotation}}
\newenvironment{Presented}{\begin{quotation} \begin{center} 
             PRESENTED AT\end{center}\bigskip 
      \begin{center}\begin{large}}{\end{large}\end{center} \end{quotation}}

\def\beq{\begin{equation}}
\def\eeq#1{\label{#1}\end{equation}}
\def\eeqn{\end{equation}}

\def\beqa{\begin{eqnarray}}
\def\eeqa#1{\label{#1}\end{eqnarray}}
\def\eeqan{\end{eqnarray}}

\let\bar=\overbar

\def\Dslash{\not{\hbox{\kern-4pt $D$}}}
\def\dslash{\not{\hbox{\kern-2pt $\del$}}}

\def\msb{{\bar{\ssstyle M \kern -1pt S}}}

\begin{document}
\begin{titlepage}
\pubblock

\vfill
\Title{Top quark transverse momentum and rapidity distributions}
\vfill
\Author{Nikolaos Kidonakis\support}
\Address{\kennesaw}
\vfill
\begin{Abstract}
I present NNLO approximate calculations, based on NNLL resummation, for top 
quark differential transverse momentum and rapidity distributions. In 
particular recent results are presented for top-pair production and 
single-top production processes and compared to the latest experimental 
data from the LHC.
\end{Abstract}
\vfill
\begin{Presented}
DPF 2013\\
The Meeting of the American Physical Society\\
Division of Particles and Fields\\
Santa Cruz, California, August 13--17, 2013\\
\end{Presented}
\vfill
\end{titlepage}
\def\thefootnote{\fnsymbol{footnote}}
\setcounter{footnote}{0}

\section{Introduction}

Differential distributions provide much more wealth of information than total cross sections and can provide better tests of the Standard Model or evidence for new physics. Top quark transverse momentum and rapidity distributions have been measured at the LHC \cite{CMS7,CMS8} and the Tevatron \cite{D0}, and more results are expected in the future.

In this contribution I present higher-order calculations of top quark transverse momentum and rapidity distributions. 
Two-loop corrections are important in top quark production (for a review see \cite{NKBP}). Approximate next-to-next-to-leading order (NNLO) corrections have 
been derived for top-antitop pair \cite{NKtop} and single-top \cite{NKstW,NKtch} production from the expansion \cite{NKNNNLO} of next-to-next-to-leading logarithm (NNLL) soft-gluon resummed double-differential cross sections \cite{NKtop,NKstW,NKtch,NKPRL}.
Here we present top-quark transverse momentum, $p_ T$, distributions 
and rapidity distributions in $t{\bar t}$ production \cite{NKtop,pnsnoweps}. 
We also present top quark $p_ T$ distributions in single-top production \cite{NKtch,pnsnoweps}.

The results from our formalism are very close to the exact NNLO for the total cross section: both the central values and the scale uncertainty are nearly the same, see the discussions in \cite{pnsnoweps}.
This was expected on theoretical grounds and past comparisons at NLO and NNLO, 
and it means that we have near-exact NNLO $p_T$ and rapidity distributions.

\section{Top pair production}

\begin{figure}[htb]
\centering
\includegraphics[height=2.3in]{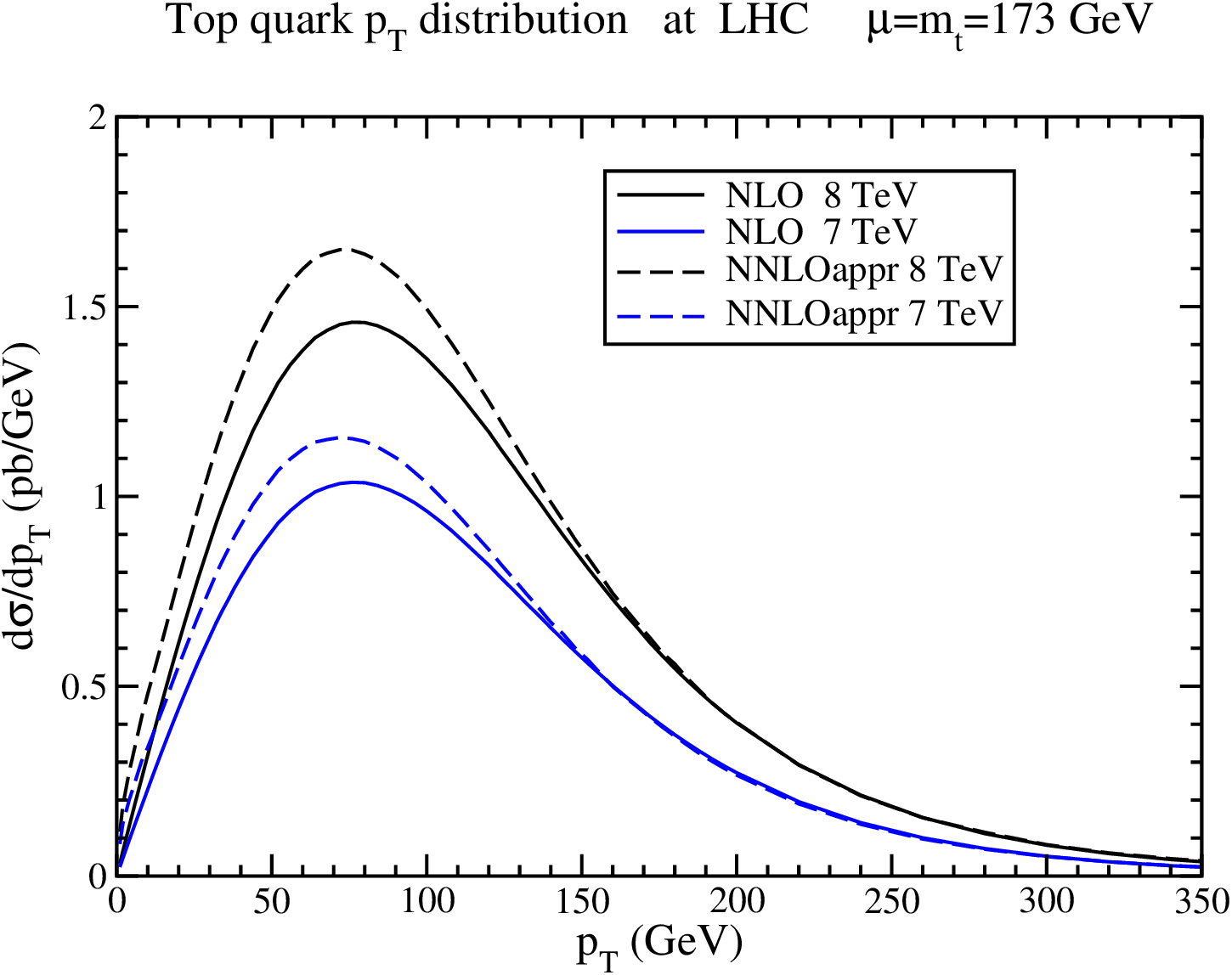}
\hspace{3mm}
\includegraphics[height=2.3in]{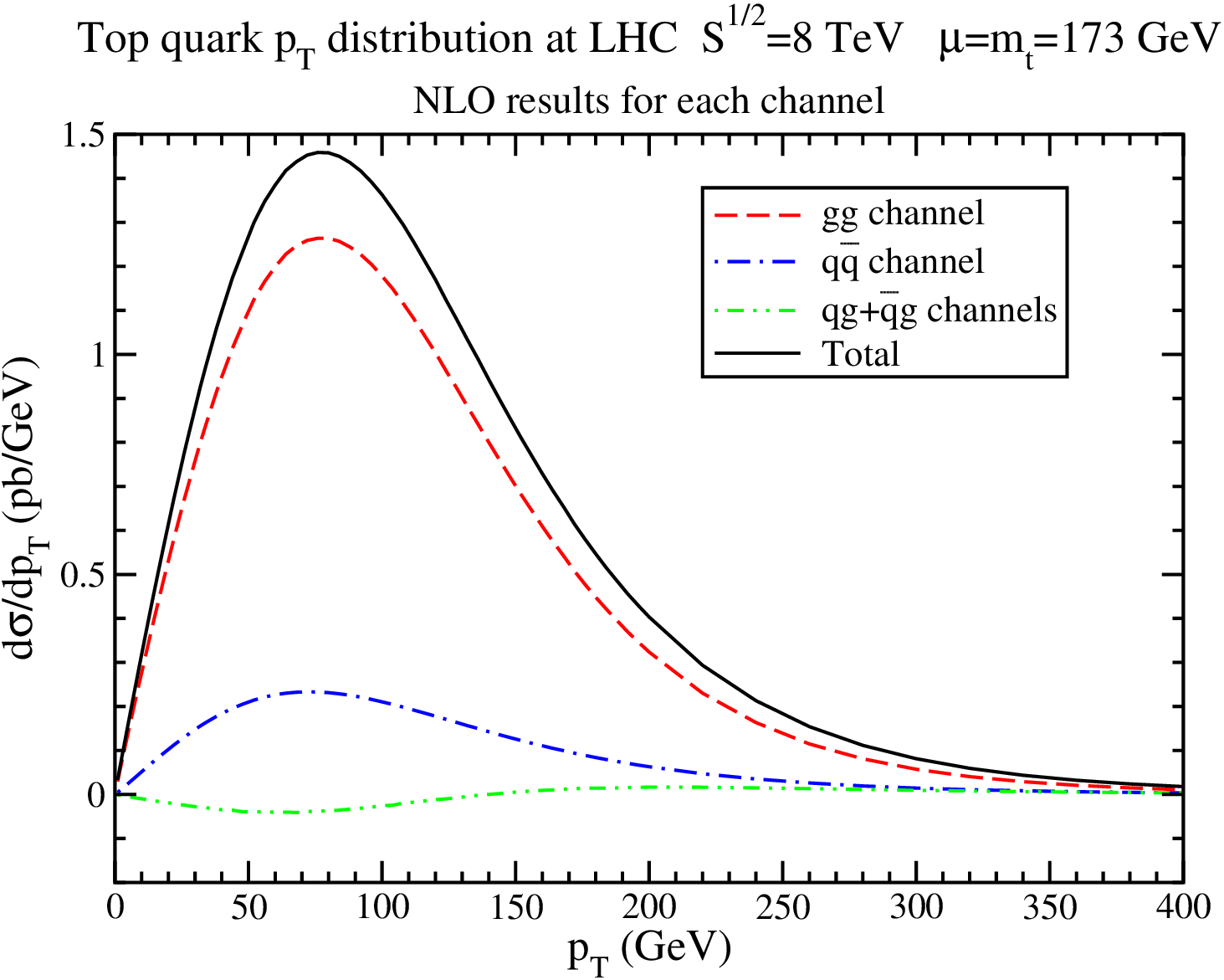}
\caption{(Left) Top quark $p_T$ distributions at LHC energies.
(Right) Contributions to the NLO top quark $p_T$ distribution from different channels at 8 TeV. }
\label{ptlhcmplot}
\end{figure}

We begin with $t{\bar t}$ production. We add NNLO soft-gluon corrections to the NLO \cite{NLOtt} differential distributions, and thus provide approximate NNLO $p_T$ and rapidity distributions \cite{NKtop}. 
In our numerical results we use MSTW2008 NNLO pdf \cite{MSTW} throughout.

In Fig. \ref{ptlhcmplot} we show the top quark $p_T$ distribution at LHC energies. On the left plot both the NLO and approximate NNLO distributions are shown at 7 TeV and 8 TeV energy, with factorization scale $\mu_F$ and renormalization scale $\mu_R$ set equal to the top quark mass.
On the right plot we show separate results for the contributions to the  NLO $p_T$ distribution from the $gg$, $q{\bar q}$, and $qg+{\bar q}g$ channels, together with the total result at 8 TeV LHC energy. Clearly the $gg$ channel is dominant.  

\begin{figure}[htb]
\centering
\includegraphics[height=2.3in]{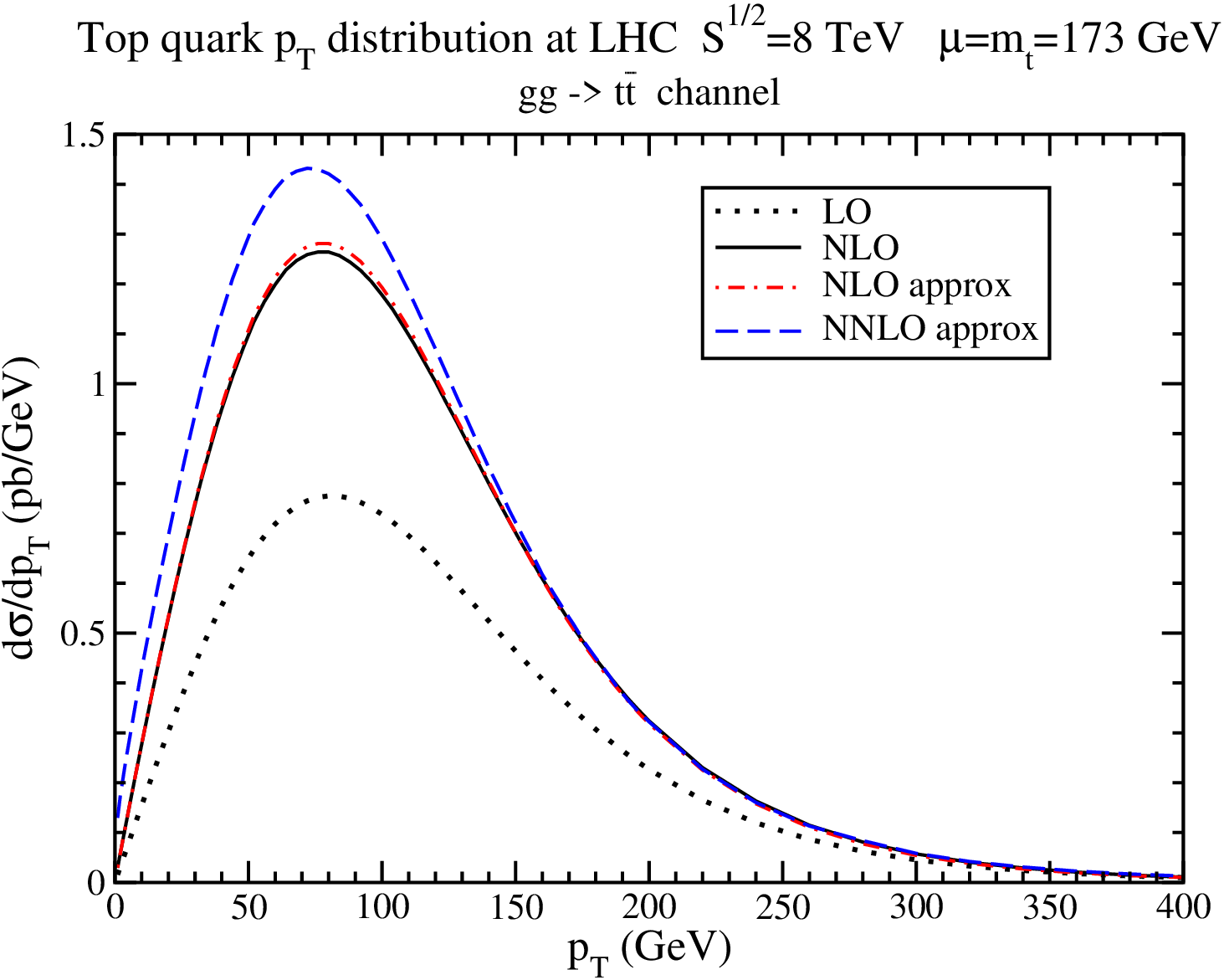}
\hspace{3mm}
\includegraphics[height=2.3in]{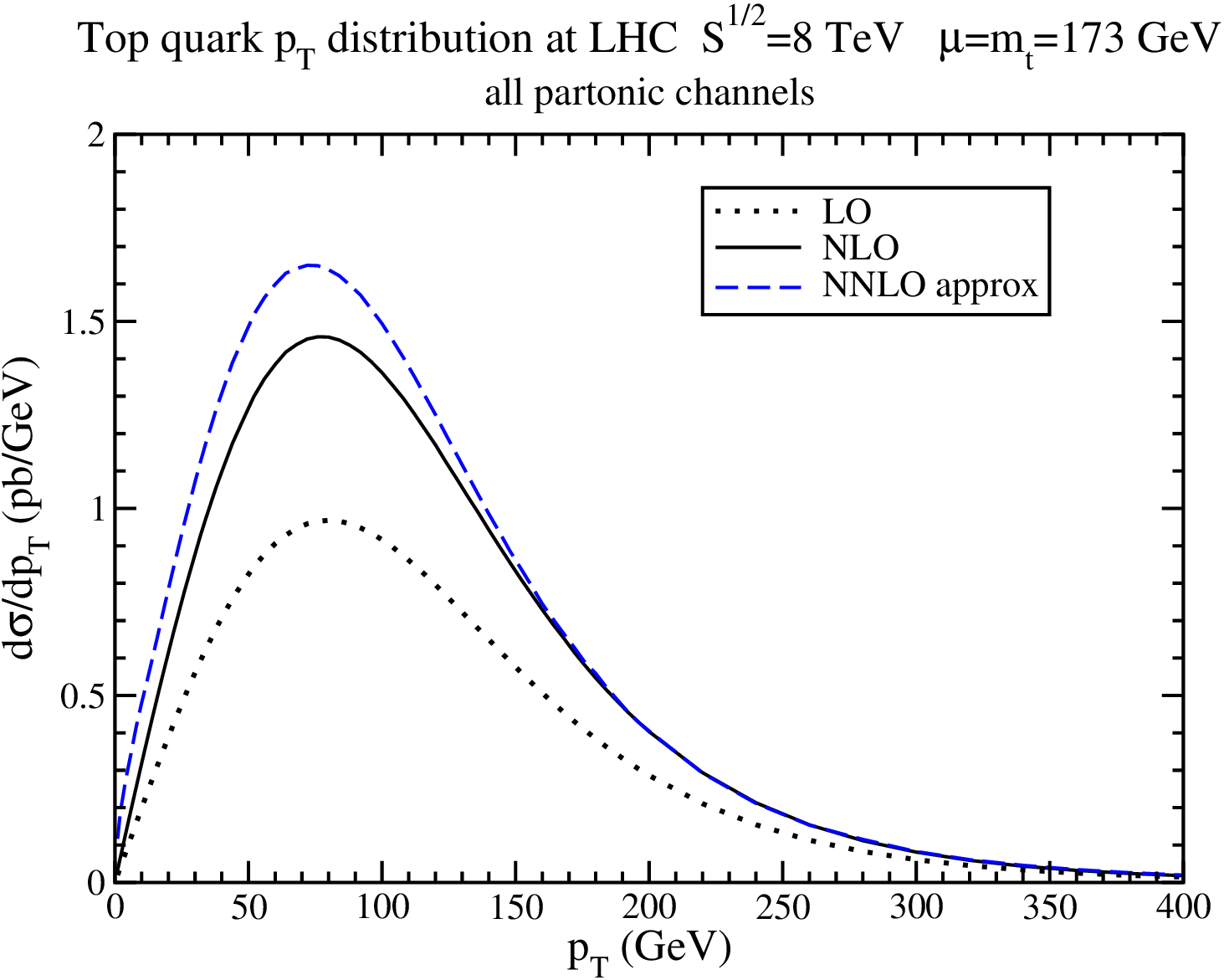}
\caption{Results at different orders for (left) the $gg$ channel contribution and (right) the total contribution to $d\sigma/dp_T$ for the top quark at 8 TeV.}
\label{pt8lhc173mplot}
\end{figure}

Figure \ref{pt8lhc173mplot} shows the LO, NLO, and approximate NNLO 
results for the top quark $p_T$ distribution at 8 TeV LHC energy. 
The left plot shows  the contribution of the $gg$ channel while the right plot shows the total distribution (sum of all channels).
It is clear that the higher-order corrections enhance the LO result. It is also clear that the approximate NLO result is almost identical to the exact NLO result. This again proves the validity and quality of the soft-gluon threshold approximation, see also the discussion in \cite{pnsnoweps}.  

\begin{figure}[htb]
\centering
\includegraphics[height=2.18in]{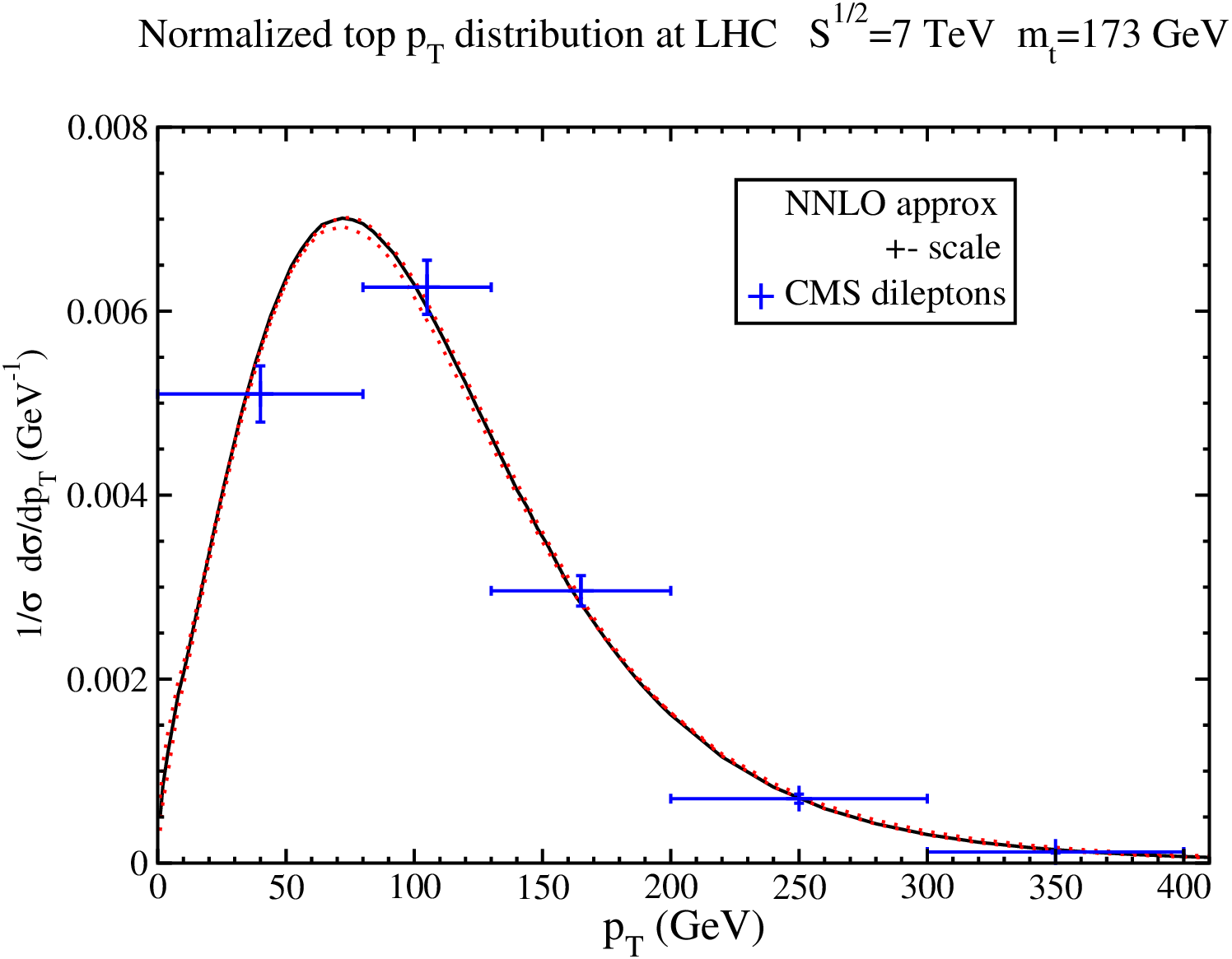}
\hspace{3mm}
\includegraphics[height=2.18in]{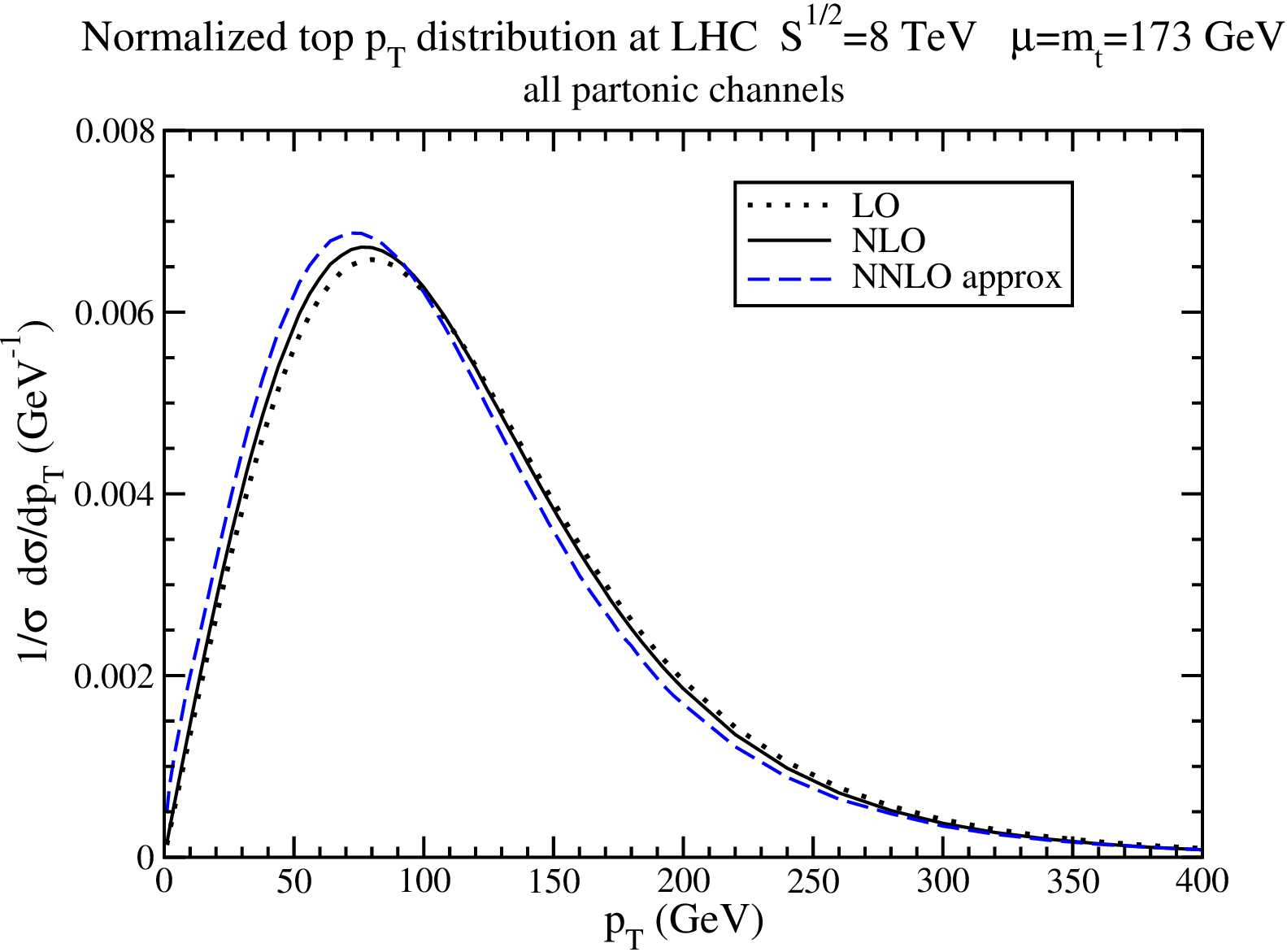}
\caption{Normalized top quark $p_T$ distributions at 7 TeV (left) and 8 TeV (right).}
\label{ptlhcnormplot}
\end{figure}

The left plot of Fig. \ref{ptlhcnormplot} shows the approximate NNLO normalized top quark $p_T$ distribution, $(1/\sigma)d\sigma/dp_T$, and compares it with recent CMS data \cite{CMS7} at 7 TeV LHC energy. The agreement with the data is excellent. At both low and high $p_T$ the NNLO approximate results provide a better description of the data than NLO or event generators (see discussion in \cite{CMS7} and \cite{pnsnoweps}). This is also true at 8 TeV LHC energy \cite{CMS8}. The change in shape is actually in accord with the trend that is already evident when going from LO to NLO. The right plot of Fig. \ref{ptlhcnormplot} shows the contributions at different orders to the normalized $p_T$ distribution at 8 TeV. As can be seen, the NLO corrections enhance the normalized $p_T$ distribution at smaller $p_T$ relative to LO, while they depress it at higher $p_T$. The NNLO approximate corrections continue this trend and they further enhance the normalized $p_T$ distribution at lower $p_T$ while further depressing it at higher $p_T$.

\begin{figure}[htb]
\centering
\includegraphics[height=2.3in]{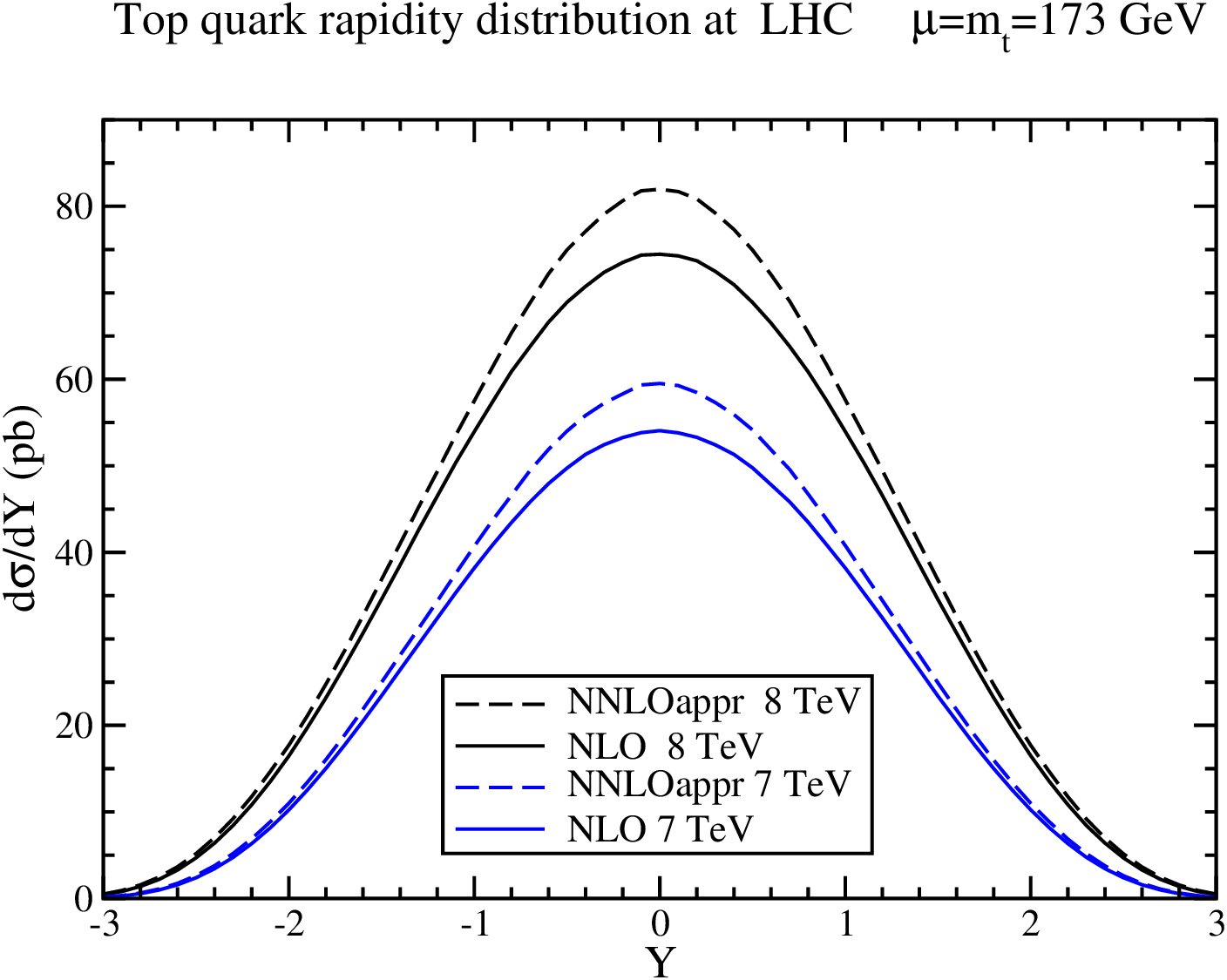}
\hspace{3mm}
\includegraphics[height=2.3in]{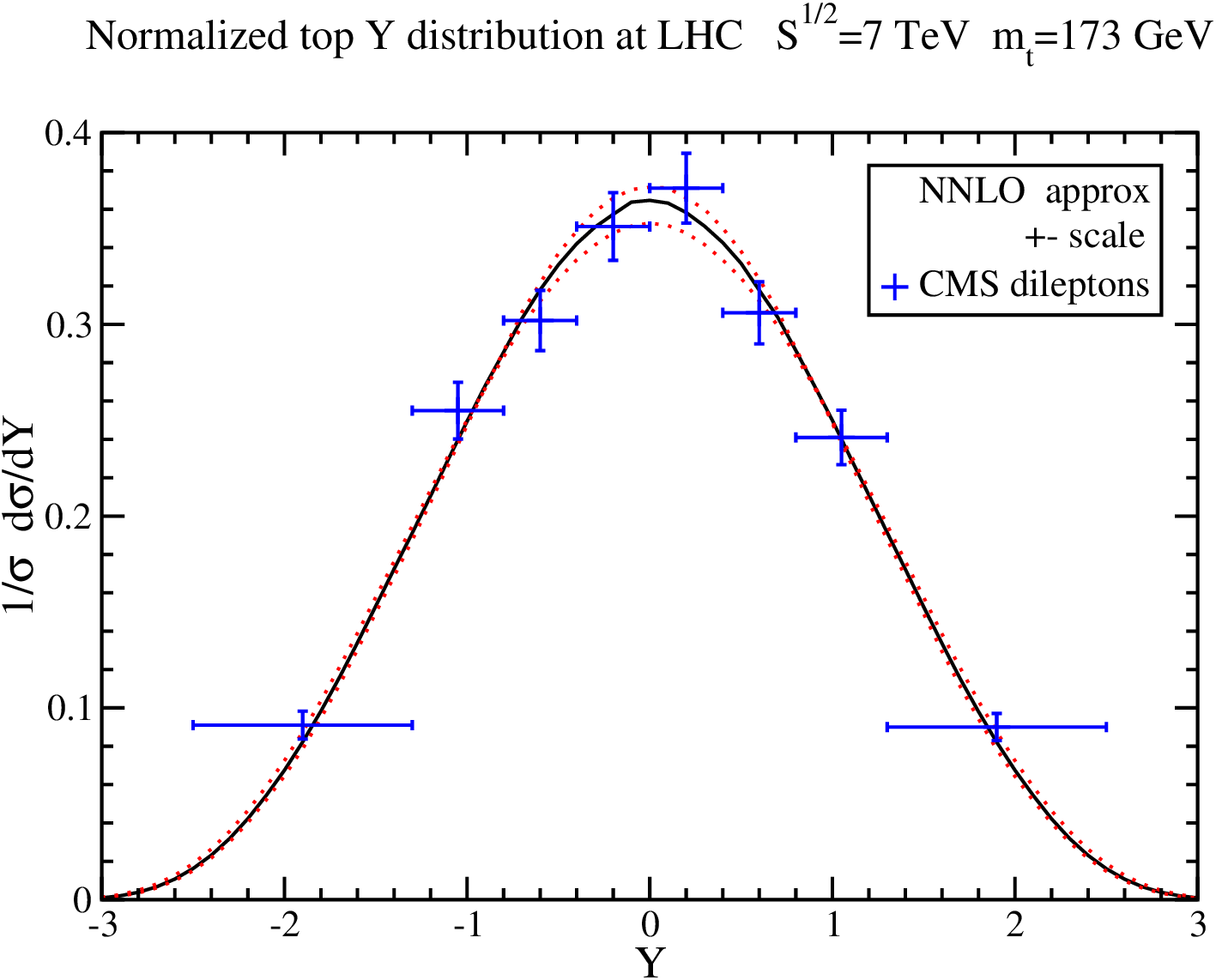}
\caption{(Left) Top quark rapidity distribution at 7 and 8 TeV LHC energies. (Right) Normalized top quark rapidity distribution at 7 TeV.}
\label{ylhcplot}
\end{figure}

Figure \ref{ylhcplot} shows the top quark rapidity distribution. The left plot displays the distribution at NLO and approximate NNLO at 7 and 8 TeV LHC energies. The right plot shows the normalized rapidity distribution at approximate NNLO together with CMS data \cite{CMS7} at 7 TeV energy. The agreement between theory and data is excellent (and similarly at 8 TeV \cite{CMS8}). 
 
\section{Single top production}

We continue with single-top production, which involves three distinct partonic channels. The numerically dominant single-top process at the LHC is $t$-channel production, and the second largest process is the associated production of a top quark with a $W$ boson. Measurements of the total cross section for $t$-channel \cite{ATLAStch,CMStch} and $tW$ \cite{ATLAStW,CMStW} production at the LHC are in very good agreement with theoretical results \cite{NKstW,NKtch}.

\begin{figure}[htb]
\centering
\includegraphics[height=2.2in]{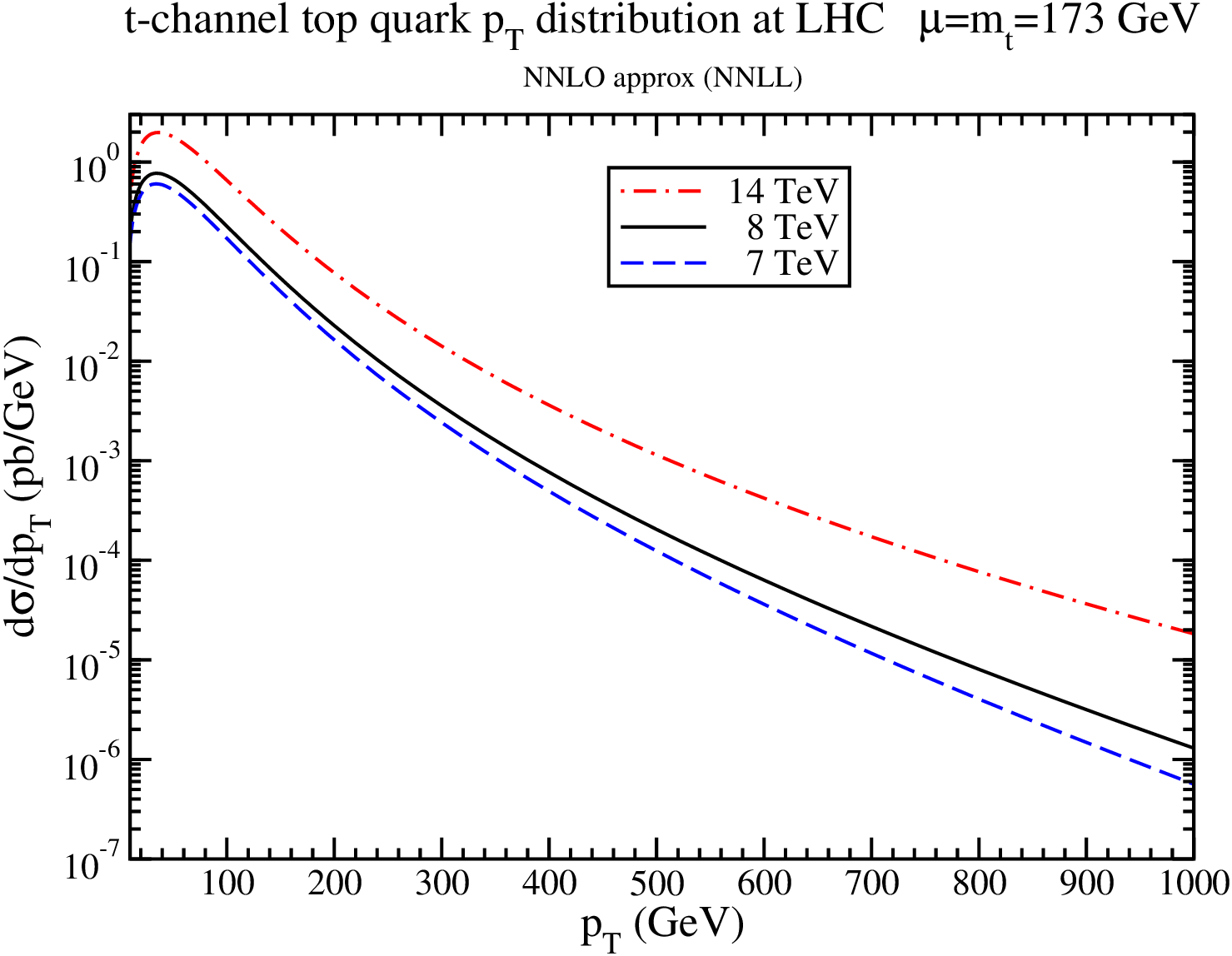}
\hspace{3mm}
\includegraphics[height=2.2in]{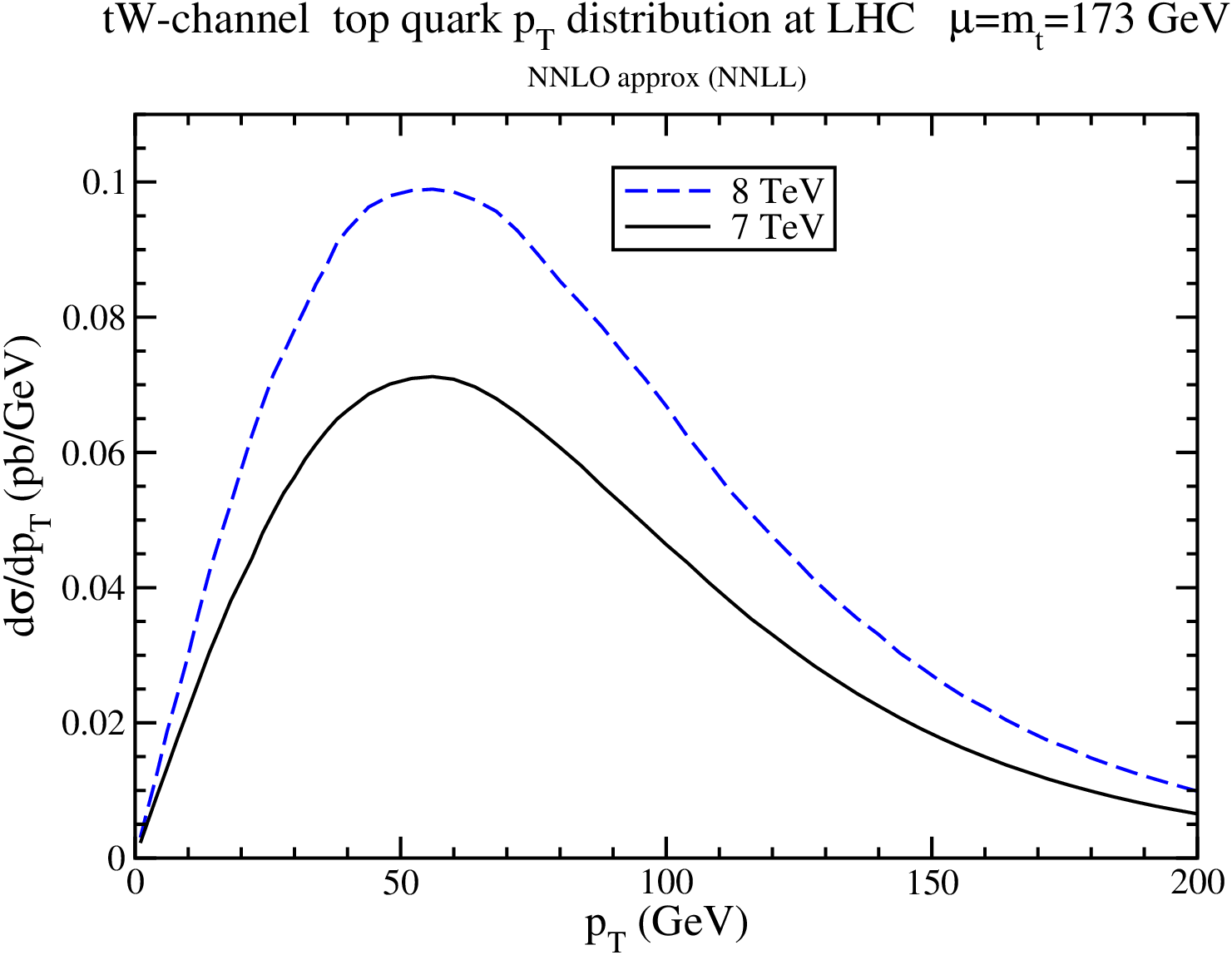}
\caption{Approximate NNLO top quark $p_T$ distributions at LHC energies in (left) $t$-channel single-top production and in (right) $tW$ production.}
\label{pttchtWplot}
\end{figure}

The NLO differential distributions for single top production were calculated in \cite{NLOsingletop}. Here we add the NNLO soft-gluon corrections from NNLL resummation to provide approximate NNLO $p_T$ distributions.
The left plot of Fig. \ref{pttchtWplot} shows the $t$-channel single top quark $p_T$ distribution at approximate NNLO at 7, 8, and 14 TeV LHC energies.
The right plot shows the $tW$-channel top quark $p_T$ distribution. 

\section{Conclusions}

The top quark differential distributions in $p_T$ and rapidity have been presented through approximate NNLO. Results for both top-pair production and single-top production are now available. The corrections are significant and in very good agreement with recent data from the LHC.

\end{document}